\documentclass[aps,prl,preprint,superscriptaddress,nofootinbib]{revtex4}

\usepackage{amsmath}
\usepackage{epsfig,bm} 

\begin{document}

\title{Isotropic Cooper Pairs with Emergent Sign Changes
 in Single-Layer Iron Superconductor}

\author{J.P. Rodriguez}

\affiliation{Department of Physics and Astronomy, 
California State University, Los Angeles, California 90032}


\begin{abstract}
We model
a single layer of heavily electron-doped FeSe
by spin-$1/2$ moments over a square lattice of iron atoms 
that include the $3d_{xz}$ and $3d_{yz}$ orbitals,
at strong on-site Coulomb repulsion.
Above half filling,
we find emergent hole bands below the Fermi level at 
the center of the one-iron Brillouin zone
in a half metal state characterized by hidden magnetic order
and by electron-type Fermi surface pockets at
wavenumbers that double the unit cell along the principal axes.
``Replicas'' of the emergent hole bands
exist at lower energy
in the two-iron Brillouin zone.
Exact calculations with two mobile electrons find evidence for isotropic
Cooper pairs that alternate in sign between the electron bands
and the emergent hole bands.
\end{abstract}

\maketitle

{\it Introduction.}
The discovery of superconductivity in iron-pnictide materials has uncovered
a new path in the search for high-temperature superconductors\cite{new_sc}.
Superconductivity has been observed recently
in a single layer of FeSe 
on a doped SrTiO$_3$ (STO) substrate\cite{xue_12,zhang_14,deng_14}
below critical temperatures as high as $100$ K \cite{ge_15}.
Electronic conduction originates from the 3$d$ orbitals of the iron atoms,
which form a square lattice.
Angle-resolved photo-emission spectroscopy(ARPES), in particular, 
reveals circular electron-type Fermi surface pockets centered at 
wave numbers
$(\pi/a){\bm{\hat x}}$  and $(\pi/a){\bm{\hat y}}$
that lie along the principal axes of the iron lattice,
where $a$ is the lattice constant\cite{liu_12,zhou_13}.
Unlike the case of most iron-pnictide materials, however,
ARPES also finds
that hole  bands centered at zero two-dimensional (2D) momentum
lie well below the Fermi level in the case of single-layer FeSe/STO.
At low temperature,
it also finds an isotropic 
gap at
the electron Fermi surface pockets\cite{peng_14,lee_14},
which is confirmed by 
scanning tunneling microscopy (STM)\cite{fan_15}.
The same set of phenomena have been recently observed 
below critical temperatures in the range $40$-$50$ K
at the surfaces
of intercalated FeSe\cite{zhao_16,niu_15,yan_15},
of alkali-metal dosed FeSe\cite{miyata_15,wen_16,ye_16,song_16},
and of voltage-gate tuned thin films of FeSe\cite{lei_16,hosono_16}.
Comparison with bulk FeSe, which has a much lower critical temperature of $8$ K,
strongly suggests that the high-temperature superconductivity exhibited above
is due to a new 2D groundstate that appears after heavy electron doping.

Calculations based on the independent-electron approximation\cite{andersen_boeri_11}
fail to describe the Fermi surfaces in single-layer FeSe/STO.
In particular, density-functional theory (DFT) typically predicts that 
the hole bands centered at zero 2D momentum 
cross the Fermi level\cite{peng_14,zhao_16,B&C}.
DFT also fails to
account for a nearby Mott insulator phase at low electron doping
in voltage-gate tuned thin films of FeSe and
in single-layer FeSe/STO\cite{zhou_14,hosono_16}.
The previous suggests that
the limit 
of strong electron-electron interactions\cite{Si&A,jpr_ehr_09}
is a better starting point to describe superconductivity 
in heavily electron-doped FeSe.

\begin{figure}
\includegraphics[scale=0.70, angle=-90]{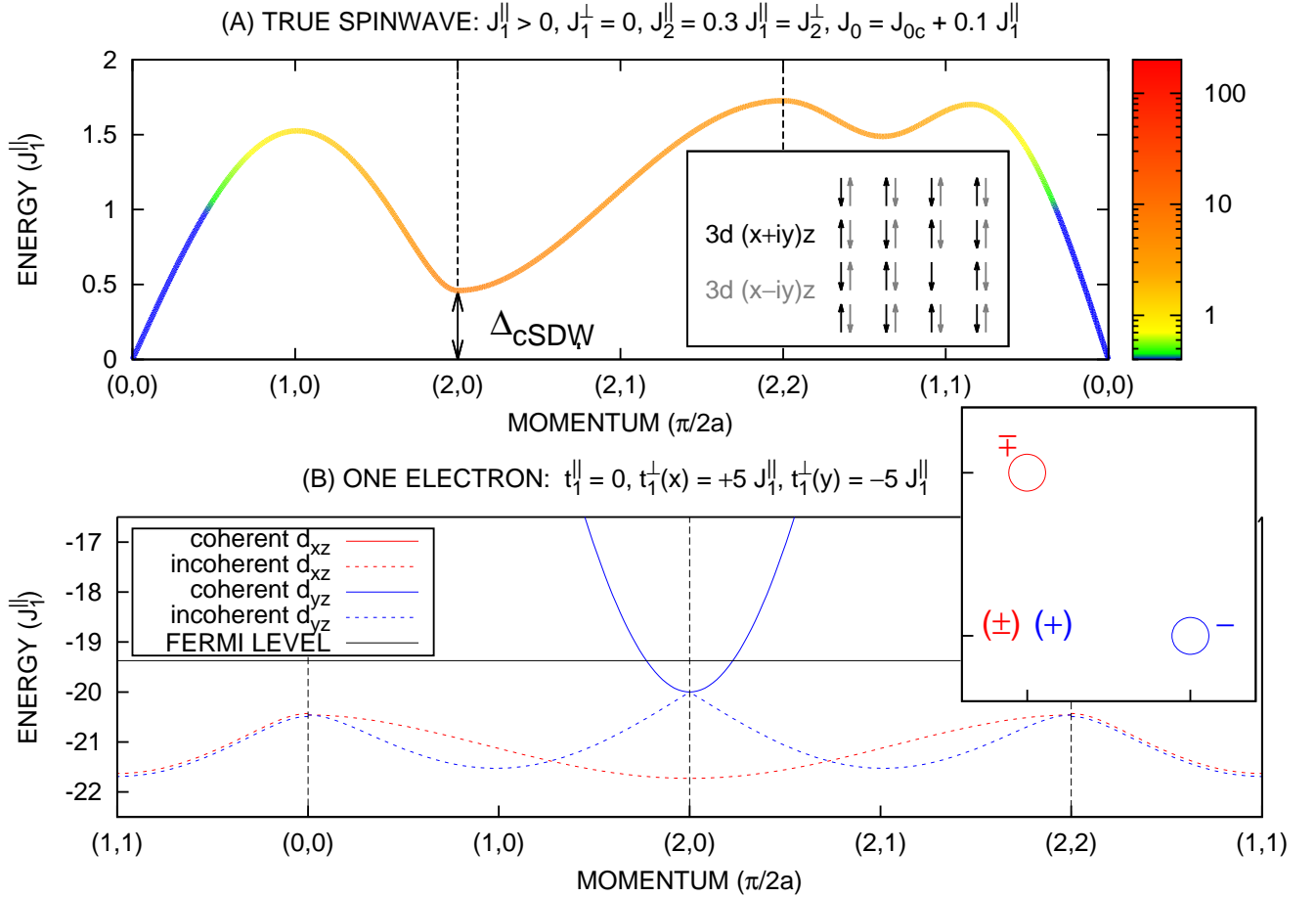}
\caption{(a) The imaginary part of the transverse spin susceptibility,
Eq. \ref{chi_perp},
in the true spin channel and
(b) the imaginary part of the one-electron propagator near half filling,
Eq. \ref{emergent_holes},
at site-orbital concentration $x=0.01$.
Not shown in (b) is intrinsic broadening due to 
the incoherent contributions in Eq. \ref{G}.}
\label{meanfield}
\end{figure}

Below, we propose that the hole bands observed by ARPES below the Fermi level 
at the Brillouin zone center
in a surface layer of FeSe
are examples of emergent phenomena.
The latter is revealed by both mean-field and exact calculations
of the one-electron spectrum
in a two-orbital $t$-$J$ model that includes only
degenerate $d_{xz}$ and $d_{yz}$ electron bands
centered at wavenumbers $(\pi/a){\bm{\hat y}}$ and $(\pi/a){\bm{\hat x}}$,
respectively,
in the one-iron Brillouin zone.
Local spin-$1/2$ moments live on $d_{(x\pm iy)z}$ orbitals, on the other hand,
which yields isotropic magnetism.
Emergent hole bands
approach the Fermi level 
at zero 2D momentum
as Hund coupling increases
inside of a half metal phase that is 
characterized by hidden N\'eel order per $d_{(x\pm iy)z}$ orbital
and by electron-type Fermi surface pockets
(inset to Fig. \ref{meanfield}b).
Emergent hole bands at wavenumber 
$(\pi/a)({\bm{\hat x}}+{\bm{\hat y}})$
in the one-iron Brillouin zone
are also predicted, but they  lie below the former ones in energy.
It is important to point out that one-electron tight-binding models
that include $d_{xz}$, $d_{yz}$, and up to $d_{xy}$ iron orbitals
are unable to account for buried hole bands at the center
{\it and} at the corner of the one-iron Brillouin zone.
(Cf. refs. \cite{raghu_08} and \cite{Lee_Wen_08}.)
Last,
exact calculations of two mobile electrons in the two-orbital $t$-$J$ model
find evidence for isotropic Cooper pairs on both
the electron pockets  and on
the emergent hole bands below the Fermi level 
as Hund coupling approaches a  quantum critical point (QCP)
at which commensurate spin-density wave (cSDW) nesting begins.
The sign of the Cooper pair wavefunction notably
alternates between the electron and hole bands\cite{mazin_08,kuroki_08}.

{\it Local Moment Model.}
Our starting point is a
two-orbital $t$-$J$ model over the square lattice,
where the on-site-orbital energy cost $U_0$
tends to infinity\cite{jpr_mana_pds_11,jpr_mana_pds_14}:
\begin{eqnarray}
H = & \sum_{\langle i,j \rangle} [-(t_1^{\alpha,\beta} {\tilde c}_{i, \alpha,s}^{\dagger} {\tilde c}_{j,\beta,s} + {\rm h.c.})
   + J_1^{\alpha,\beta} {\bm S}_{i, \alpha} \cdot {\bm S}_{j, \beta}]
   +  \sum_{\langle\langle i,j \rangle\rangle} J_2^{\alpha,\beta}
      {\bm S}_{i, \alpha} \cdot {\bm S}_{j, \beta} \nonumber \\
& + \sum_i (J_0 {\bm S}_{i, d-}\cdot {\bm S}_{i, d+} 
 +  U_0^{\prime} {\bar n}_{i,d+} {\bar n}_{i,d-} \bigr).
\label{tJ}
\end{eqnarray}
Above, ${\bm S}_{i,\alpha}$ is the spin operator that acts on spin $s_0 = 1/2$ states
of $d- = d_{(x-iy)z}$ and $d+ = d_{(x+iy)z}$ orbitals $\alpha$
in iron atoms at sites $i$.  
Repeated orbital and spin indices
in the hopping and Heisenberg exchange terms above
are summed over.
Nearest neighbor and next-nearest neighbor Heisenberg exchange
across the links $\langle i,j\rangle$ and $\langle\langle i,j\rangle\rangle$
is controlled by 
exchange coupling constants
$J_1^{\alpha,\beta}$ and $J_2^{\alpha,\beta}$, respectively.  
Hopping of an electron in orbital $\alpha$ to a
nearest-neighbor orbital $\beta$ is controlled by
the matrix element $t_1^{\alpha,\beta}$.
We adopt the Schwinger-boson ($b$) slave-fermion ($f$) representation 
for the creation operator of the correlated electron\cite{Arovas_Auerbach_88,kane_89,Auerbach_Larson_91}
at or {\it above} half filling:
${\tilde c}_{i,\alpha,s}^{\dagger} = f_{i,\alpha}^{\dagger} b_{i,\alpha,s}$
with the constraint
%
\begin{equation}
2 s_0 = b_{i,\alpha,\uparrow}^{\dagger} b_{i,\alpha,\uparrow} 
+ b_{i,\alpha,\downarrow}^{\dagger} b_{i,\alpha,\downarrow} 
+ f_{i,\alpha}^{\dagger} f_{i,\alpha}
\label{constraint}
\end{equation}
enforced at each site-orbital to impose the $U_0\rightarrow\infty$ limit
on electrons with spin $s_0 = 1/2$.  
Finally, $J_0$ is a ferromagnetic exchange coupling constant
that imposes Hund's Rule,
while the last term in (\ref{tJ}) represents the additional energy cost
of a fully occupied iron atom.
Here 
${\bar n}_{i,\alpha} = \sum_s {\tilde c}_{i,\alpha,s}^{\dagger}{\tilde c}_{i,\alpha,s} - 1$
counts singlet pairs at site-orbitals.
Last,
notice that
$d\pm\rightarrow e^{\pm i\theta} d\pm$ is equivalent to
a rotation of the orbitals by an angle $\theta$ about the $z$ axis.
Spin and occupation operators remain invariant under it.
Magnetism described by 
the two-orbital $t$-$J$ model (\ref{tJ}) is hence isotropic,
which suppresses orbital order.


Semi-classical calculations of 
the Heisenberg model that corresponds to (\ref{tJ}) 
at half filling
find a QCP
that separates a cSDW at strong Hund coupling
from a hidden antiferromagnet
at weak Hund coupling 
when diagonal frustration is present\cite{jpr_10}: e.g.
$J_1^{\parallel} > 0$, $J_1^{\perp} = 0$, and  $J_2^{\parallel} = J_2^{\perp} > 0$.
Here,
$\parallel$ and $\perp$ represent intra-orbital ($d\pm d\pm$)
and inter-orbital ($d\pm d\mp$) superscripts.
The hidden-order magnet shows N\'eel spin order per $d\pm$  orbital
following the inset to Fig. \ref{meanfield}a.
Ideal hopping of electrons within an antiferromagnetic sublattice,
$t_1^{\parallel} = 0$ and
$t_1^{\perp}({\bm{\hat x}}) = -t_1^{\perp}({\bm{\hat y}}) > 0$,
leaves such hidden magnetic order intact in the semi-classical limit,
$s_0\rightarrow \infty$.  
Below, we employ a
mean-field approximation of (\ref{tJ}) and (\ref{constraint}) 
to study this state near the QCP.
It reveals a half metal 
with circular Fermi surface pockets
at wavenumbers $(\pi/a){\bm{\hat x}}$ and $(\pi/a){\bm{\hat y}}$,
for electrons in the $d_{yz}$ orbital
and $d_{xz}$ orbital, respectively.

{\it Spin-Fluctuations, One-Electron Spectrum.}
Following Arovas and Auerbach\cite{Arovas_Auerbach_88}, 
we first rotate the spins quantized along the $z$ axis
on {\it one} of the antiferromagnetic sublattices
shown in the inset to Fig. \ref{meanfield}a
by an angle $\pi$ about the $y$ axis.
This decouples the up and down spins between the two sublattices\cite{sup_mat}.
We next define mean fields that are set by the pattern of antiferromagnetic 
versus ferromagnetic  pairs of neighboring spins\cite{Arovas_Auerbach_88}
in the hidden magnetic order:
$Q_0 = \langle b_{i,d-,s} b_{i,d+,s}\rangle$,
$Q_{1}^{\parallel} = \langle b_{i,d\pm,s} b_{j,d\pm,s}\rangle$ and
$Q_{2}^{\perp} = \langle b_{i,d\pm,s} b_{j,d\mp,s}\rangle$
on the antiferromagnetic links
versus
$Q_{1}^{\perp} = \langle b_{i,d\pm,s}^{\dagger} b_{j,d\mp,s}\rangle$ and
$Q_{2}^{\parallel} = \langle b_{i,d\pm,s}^{\dagger} b_{j,d\pm,s}\rangle$
on the ferromagnetic links of the hidden N\'eel state.
Subscripts $0$, $1$ and $2$ represent on-site, nearest neighbor and
next-nearest neighbor links.
We add to that list the mean field 
$P_1^{\perp} = {1\over 2} \langle f_{i,d\pm}^{\dagger} f_{j,d\mp}\rangle$
for nearest-neighbor hopping of electrons across the two orbitals.  
It has $d$-wave symmetry.
The corresponding mean-field approximation for 
the $t$-$J$ model Hamiltonian (\ref{tJ}) then has the form
$H_b + H_f$, 
where 
%
$$H_b = {1\over 2}\sum_{k}\sum_{s} \{ 
\Omega_{fm}(k) [b_s^{\dagger}(k) b_s(k) +  b_s(-k) b_s^{\dagger}(-k)]
            + \Omega_{afm}(k) [b_s^{\dagger}(k) b_s^{\dagger}(-k) + b_s(-k) b_s(k)]\}$$
is the Hamiltonian for free Schwinger bosons,
and where
$H_f = \sum_k \varepsilon_f(k) f^{\dagger}(k) f(k)$ 
is the Hamiltonian for free slave fermions.
Here, 
$k = (k_0,{\bm k})$ 
is the  3-momentum for these excitations,
where 
the quantum numbers $k_0 = 0$ and $\pi$
represent even and odd superpositions of the $d-$ and $d+$ orbitals:
$d_{xz}$ and $(-i)d_{yz}$.

Enforcing the infinite-$U_0$ constraint (\ref{constraint})
on average over the bulk then results 
in ideal Bose-Einstein condensation (BEC) of the Schwinger bosons
into degenerate groundstates at
$k = 0$ and $(\pi,\pi/a,\pi/a)$
in the zero-temperature limit:
$\langle b_{i,d\pm,s}\rangle = s_0^{1/2}$ at large $s_0$.
(See Fig. \ref{meanfield}a and supplemental Fig. S1.)
In such case,
all five mean fields among the Schwinger bosons therefore
take on the unique value $Q = s_0$ \cite{sup_mat}.
This results in diagonal and off-diagonal Hamiltonian matrix elements
\begin{eqnarray*}
\Omega_{fm}(k) &=& 
(1-x)^2 s_0 (J_0 + 4 J_1^{\parallel} + 4 J_2^{\perp} \\
&& - 4 J_1^{\prime\prime\perp}[1-e^{ik_0}\gamma_{1+}({\bm k})]
- 4 J_2^{\parallel} [1-\gamma_2({\bm k})])\\
\Omega_{afm}(k) &=& 
 -(1-x)^2 s_0 [J_0 e^{i k_0} 
+4 J_1^{\parallel}  \gamma_{1+}({\bm k})
+4 J_2^{\perp} e^{ik_0}\gamma_2({\bm k})]
\end{eqnarray*}
for free Schwinger bosons,
and the energy eigenvalues
$\varepsilon_f(k) = -8 s_0 t_{1}^{\perp}({\bm{\hat x}}) e^{ik_0} \gamma_{1-}({\bm k})$
for free slave fermions.
Above,
$J_1^{\prime\prime\perp} = J_1^{\perp} - 2 t_{1}^{\perp}({\bm{\hat x}}) P_{1}^{\perp}({\bm{\hat x}}) /(1-x)^2 s_0$,
while
$\gamma_{1\pm} ({\bm k}) = {1\over 2} ({\rm cos}\, k_x a\, \pm \,{\rm cos}\, k_y a)$
and $\gamma_2 ({\bm k}) = {1\over 2} ({\rm cos}\, k_+ a\,+\,{\rm cos}\, k_- a)$,
with $k_{\pm} = k_x \pm k_y$.
Slave fermions 
in $d_{xz}$ and $d_{yz}$ orbitals
lie within circular Fermi surfaces centered at wavenumbers
$(\pi/a){\bm{\hat y}}$ and $(\pi/a){\bm{\hat x}}$, respectively, 
with Fermi wave vector $k_F a = (4\pi x)^{1/2}$
at low electron doping per iron orbital,
$x \ll 1$.
(See the inset to Fig. \ref{meanfield}b.)
The mean inter-orbital electron hopping amplitude is then approximately
$P_{1}^{\perp}({\bm{\hat x}}) = x/2$.

The dynamical spin correlation function
$\langle S_y S_y^{\prime}\rangle$
is obtained directly from 
the above Schwinger-boson-slave-fermion mean field theory.  
It is given by an Auerbach-Arovas expression 
at non-zero temperature
that  is easily evaluated in 
the zero-temperature limit \cite{jpr_mana_pds_14,Auerbach_Arovas_88},
where ideal BEC
of the Schwinger bosons into the 
degenerate groundstates at 3-momenta
$k = 0$ and $(\pi,\pi/a,\pi/a)$ occurs.
It is one half the transverse spin correlator,
which under ideal BEC and at large $s_0$ reads 
\begin{equation}
i \langle S^{(+)} S^{\prime (-)} \rangle |_{k,\omega}
= (1-x)^2 s_0 (\Omega_{+}/\Omega_{-})^{1/2} 
([\omega_b(k) - \omega ]^{-1} + [\omega_b(k) + \omega ]^{-1}).
\label{chi_perp}
\end{equation}
Here,
$\omega_b = (\Omega_{fm}^2 - \Omega_{afm}^2)^{1/2}$
is the energy dispersion of the Schwinger bosons,
and  $\Omega_{\pm} = \Omega_{fm}\pm\Omega_{afm}$.
Figure \ref{meanfield}a depicts 
the imaginary part of the 
transverse susceptibility (\ref{chi_perp})
in the true spin channel,
$k_0 = 0$,
at sub-critical Hund coupling.
It reveals
a spin gap 
at cSDW wave numbers
$(\pi/a) {\bm{\hat x}}$ and $(\pi/a) {\bm{\hat y}}$
of the form
$\Delta_{cSDW} = (1-x)^2 (2 s_0) (4 J_2^{\perp} - J_{0c})^{1/2} {\rm Re}\, (J_0 - J_{0c})^{1/2}$.
Here,
$- J_{0c} = 2 (J_1^{\parallel} - J_1^{\perp}) - 4 J_2^{\parallel}
  + (1-x)^{-2} s_0^{-1} 2  t_{1}^{\perp}({\bm{\hat x}}) x$
is the critical Hund coupling at which $\Delta_{cSDW}\rightarrow 0$.
Notice that inter-orbital hopping
stabilizes the hidden half metal state.
The autocorrelator of the hidden spin
${\bm S}_{i,d-} - {\bm S}_{i,d+}$,
(\ref{chi_perp}) at $k_0 = \pi$,
also shows the above spin gap
 at cSDW momenta, $\Delta_{cSDW}$,
in addition to a hidden-order Goldstone mode
at N\'eel wavenumber $(\pi/a)({\bm{\hat x}}+{\bm{\hat y}})$\cite{sup_mat}.

The electronic structure of the hidden half metal state 
can also be obtained directly from the above
Schwinger-boson-slave-fermion mean field theory.  
In  particular, the one-electron propagator
is given by the convolution
of the conjugate propagator for Schwinger bosons with 
the propagator for slave fermions 
in 3-momentum and in frequency.
A summation of Matsubara frequencies yields the expression\cite{sup_mat}
\begin{eqnarray}
G(k,\omega) = {1\over{\cal N}}\sum_q &\Bigl [\Bigl({1\over 2} {\Omega_{fm}\over{\omega_b}}\Bigl|_{q-k} + {1\over 2}\Bigr)
{{n_B[\omega_b (q-k)] + n_F[\varepsilon_f (q)-\mu]}\over{\omega + \omega_b(q-k) - \varepsilon_f(q)+\mu}}  \nonumber  \\
& +\Bigl({1\over 2} {\Omega_{fm}\over{\omega_b}}\Bigl|_{q-k} - {1\over 2}\Bigr)
{{n_B[\omega_b (q-k)] + n_F[\mu-\varepsilon_f(q)]}\over{\omega - \omega_b(q-k) - \varepsilon_f(q)+\mu}}\Bigr] . 
\label{G}
\end{eqnarray}
Above, 
$n_B$ and $n_F$ denote the Bose-Einstein and the Fermi-Dirac distributions,
and $\mu$ denotes the chemical potential of the slave fermions.
Ideal BEC
of the Schwinger bosons at 3-momenta
$q-k = 0$ and $(\pi,\pi/a,\pi/a)$
results in the following
coherent contribution to the electronic spectral function
at zero temperature and at large $s_0$:
${\rm Im}\, G_{\rm coh}(k,\omega) = s_0 
{\pi} \delta[\omega + \mu - \varepsilon_f(k)]$.
It reveals degenerate electron bands
for $d_{xz}$ and $d_{yz}$ orbitals
centered at cSDW wave numbers
${\bm Q}_{0} =(\pi/a){\bm{\hat y}}$ and 
${\bm Q}_{\pi} = (\pi/a){\bm{\hat x}}$ , respectively.
The electron Fermi surface pockets at $\omega = 0$
are depicted by the inset to Fig. \ref{meanfield}b.
At energies 
below the Fermi level, $\omega < 0$,
the remaining contribution 
is exclusively due to
the first fermion term in (\ref{G}).
Inspection of Fig. \ref{meanfield}b (solid lines)
yields the following expression for it
in the limit near half-filling,
$k_F a\rightarrow 0$,
at large $t/J$ \cite{diverge}:
%
\begin{equation}
{\rm Im}\, G_{\rm inc}(k,\omega) \cong \sum_{q_0=0,\pi}
{\pi\over 2} x \Biggl[{1\over 2}
+{1\over 2}{\Omega_{fm}\over{\omega_b}}\Bigl|_{(q_0-k_0,{\bm Q}_{q_0}-{\bm k})}\Biggr]
\delta[\omega + \epsilon_F + \omega_b(q_0-k_0,{\bm Q}_{q_0} - {\bm k})].
\label{emergent_holes}
\end{equation}
Figure \ref{meanfield}b displays the emergent hole bands
predicted above.
They lie $\epsilon_F + \Delta_{cSDW}$ below the Fermi level,
with degenerate maxima at ${\bm k} = 0$ and $(\pi/a)({\bm{\hat x}}+{\bm{\hat y}})$.
Here, $\epsilon_F = (2 s_0) t_1^{\perp}({\bm{\hat x}})(k_F a)^2$
is the Fermi energy.
The emergent hole bands also show intrinsic broadening in frequency
at zero temperature,
which makes them incoherent.
Outside the critical region, at large $t/J$, 
the broadening is
$\Delta\omega\sim k_F|{\bm\nabla} \omega_b|_{{\bm Q}-{\bm k}}$.
It remains small at the previous maxima\cite{broadening}.
Last, the emergent hole bands predicted by (\ref{emergent_holes})
are anisotropic:
e.g., the $d_{yz}$ hole band at zero 2D momentum
has mass anisotropy $|m_x| < |m_y|$. (Cf. ref. \cite{t1_parallel}.)

\begin{figure}
\includegraphics[scale=0.70, angle=-90]{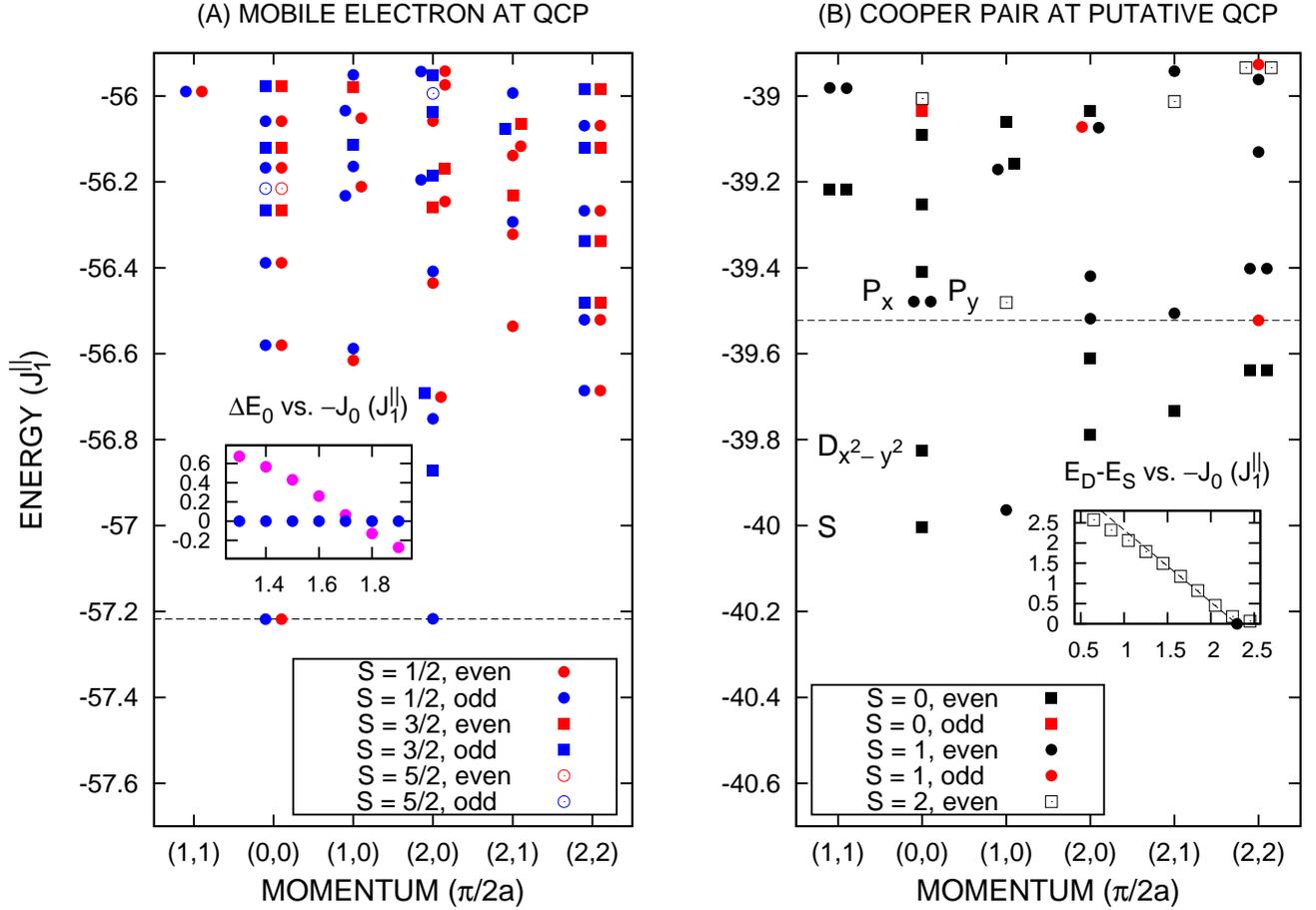}
\caption{(a) Low-energy spectrum of two-orbital $t$-$J$ model,
Eq. (\ref{tJ}) plus constant ${3\over 4}(N_{\rm Fe}-1)J_0$, over a $4\times 4$ lattice,
with one electron more than half filling.  Model parameters coincide with those listed
by Fig. \ref{meanfield}, except $t_1^{\parallel} = 2\, J_1^{\parallel}$
and $-J_0 = 1.733 \, J_1^{\parallel}$.
(b) Low-energy spectrum of Eq. (\ref{tJ}) plus repulsive interactions (see text)
plus constant ${1\over 4}(N_{\rm Fe}-2)J_0$,
but with two electrons more than half filling, with $-J_0 = 2.25 \, J_1^{\parallel}$,
and with $U_0^{\prime} = {1\over 4}J_0 + 1000\, J_1^{\parallel}$.
Some points in spectra are artificially
moved slightly off their quantized values along the momentum axis
for the sake of clarity.}
\label{exact}
\end{figure}

Adding intra-orbital electron hopping, $t_1^{\parallel} > 0$,
brings the emergent hole bands at
wavenumber $(\pi/a)({\bm{\hat x}}+{\bm{\hat y}})$
down in energy below the ones at zero 2D momentum.
This is confirmed by exact calculations of
the two-orbital $t$-$J$ model
with one electron more than half filling
over a $4\times 4$ lattice of iron atoms under periodic boundary conditions.
The previous Schwinger-boson-slave-fermion description (\ref{constraint})
for spin $s_0 = 1/2$ electrons is exploited to
impose strong on-site-orbital Coulomb repulsion.
Details are given in ref. \cite{jpr_mana_pds_14}.
Figure \ref{exact}a shows the exact spectrum at the QCP,
where $\Delta_{cSDW}\rightarrow 0$.  
The $t$-$J$ model parameters coincide
with those set by Fig. \ref{meanfield}, 
but with $t_1^{\parallel} = 2\, J_1^{\parallel}$,
and with Hund coupling tuned to the critical value
$-J_0 = 1.733\, J_1^{\parallel}$.
Red states have even parity under orbital swap, $P_{d,{\bar d}}$,
while blue states have odd parity under it.
Notice that the lowest-energy
doubly-degenerate states
at wave number $(\pi/a)({\bm{\hat x}}+{\bm{\hat y}})$,
which are spin-$1/2$,
lie $0.5\,J_1^{\parallel}$ in energy 
above the doubly-degenerate spin-$1/2$ groundstates at zero 2D momentum.
The latter states (purple) move up in energy off the Fermi level
set by the groundstates at cSDW momenta
as Hund coupling falls below the critical value,  
and they become nearly degenerate with the former states
in the absence of Hund's Rule.
This dependence on Hund coupling is demonstrated 
by the inset to Fig. \ref{exact}a
and by supplemental Fig. S3.
The exact low-energy spectrum at sub-critical Hund coupling
is therefore consistent with the emergent hole bands obtained
by the meanfield approximation, Fig. \ref{meanfield}b, 
but with the hole bands centered at wavenumber 
$(\pi/a)({\bm{\hat x}}+{\bm{\hat y}})$ 
pulled down to lower energy.
Last, Fig. \ref{exact}a shows that
the even parity ($d_{xz}$) and odd parity ($d_{yz}$)
spin-$1/2$ groundstates at wavenumber
$(\pi/2a){\bm{\hat x}}$ are nearly degenerate,
which suggests isotropic emergent hole bands
at zero 2D momentum near the QCP.

{\it Cooper Pairs.}
Figure \ref{exact}b shows the spectrum of
the same two-orbital $t$-$J$ model (\ref{tJ}),
but with two electrons more than half filling.
A {\it repulsive} interaction
has been added to the Heisenberg exchange terms
in order to reduce finite-size effects:
${\bm S}_{i, \alpha} \cdot {\bm S}_{j, \beta}\rightarrow
{\bm S}_{i, \alpha} \cdot {\bm S}_{j, \beta}+{1\over 4}n_{i, \alpha} n_{j, \beta}$,
equal to $1/2$ the spin-exchange operator.
Here, $n_{i, \alpha}$ counts the net occupation of {\it holes} per site-orbital.
Also,
the on-site repulsion between mobile electrons
in the $d+$ and $d-$ orbitals, respectively,
is set to a large value
$U_0^{\prime} = {1\over 4}J_0 + 1000\, J_1^{\parallel}$.
The Schwinger-boson-slave-fermion description of the correlation electron
(\ref{constraint}) is  again employed, with $s_0 = 1/2$.
Details are given in ref. \cite{jpr_16}.
Last, the ferromagnetic Hund's Rule exchange coupling constant
is  tuned to the critical value $J_{0} = - 2.25\, J_1^{\parallel}$,
at which $\Delta_{cSDW}\rightarrow 0$.
This is depicted by the dashed horizontal line in Fig. \ref{exact}b,
which shows the degeneracy between the cSDW spin resonance
at wavenumber $(\pi/a){\bm{\hat x}}$
with the hidden-order spin resonance
at wavenumber $(\pi/a)({\bm{\hat x}}+{\bm{\hat y}})$.
The  former is even (black)
under swap of the orbitals,
$d- \leftrightarrow d+$,
while latter is odd (red) under it.
Notice that the groundstate and the second excited state 
both lie under a continuum of states at zero net momentum.  
They respectively have even and odd parity
under a reflection about the $x$-$y$ diagonal.
We therefore assign $S$ symmetry to the groundstate bound pair
and $D_{x^2-y^2}$ symmetry to the excited-state bound pair.
The dependence of the energy-splitting between these two states on Hund coupling
is shown by the inset to Fig. \ref{exact}b.  
It provides evidence for a true QCP 
in the thermodynamic limit
at $-J_0 = 2.30 \, J_1^{\parallel}$,
where the $s$-wave and $d$-wave bound states become degenerate.

\begin{figure}
\includegraphics[scale=0.70, angle=0]{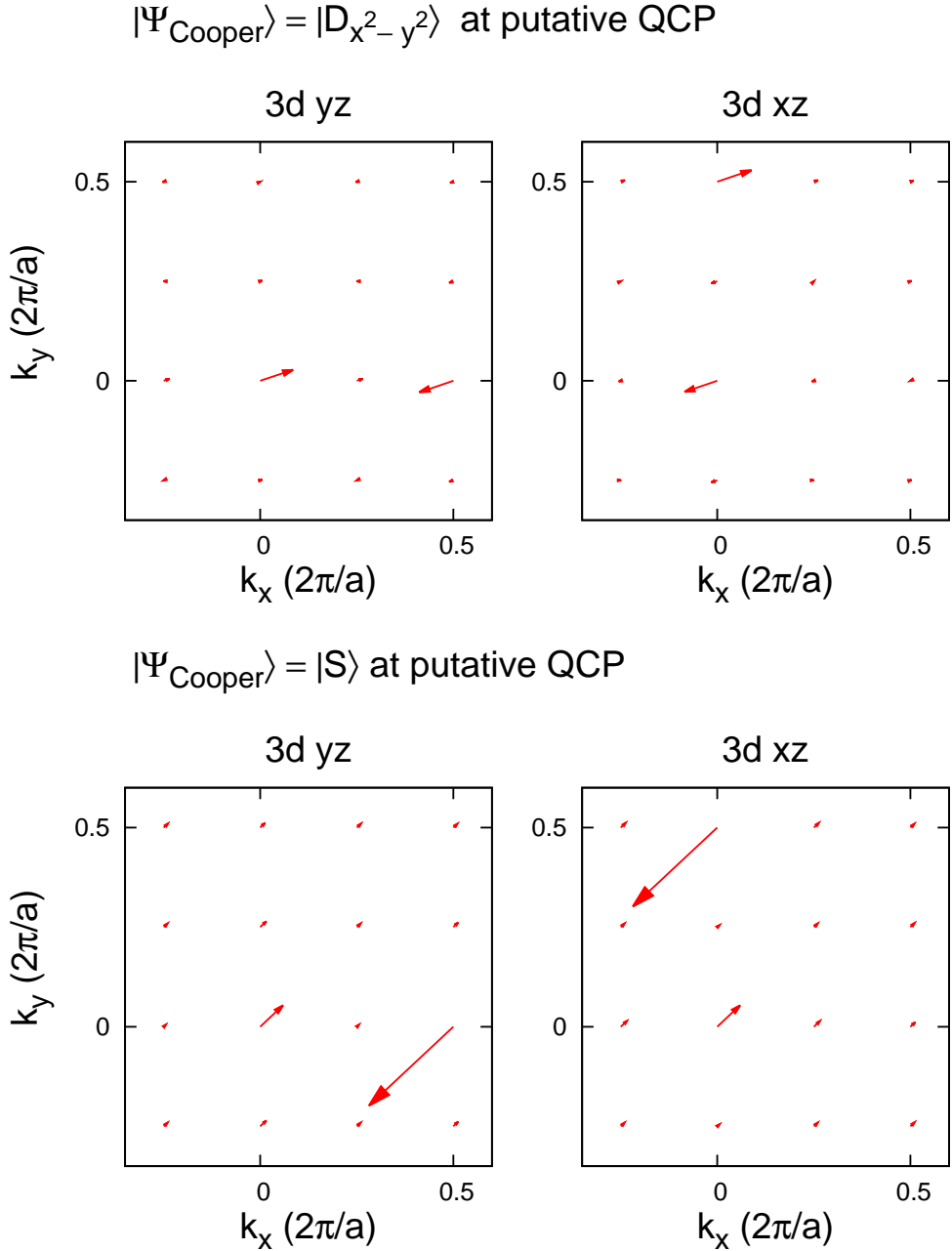}
\caption{The complex order parameter for superconductivity,
Eq. \ref{iF},
symmetrized with respect to both reflections about the principal axes.}
\label{s+-d+-}
\end{figure}

Figure \ref{s+-d+-} depicts
the order parameters for superconductivity
of the two bound pair states shown in Fig. \ref{exact}b:
\begin{equation}
iF(k_0,{\bm k}) = \langle \Psi_{\rm Mott}|
{\tilde c}_{\uparrow}(k_0,{\bm k})
{\tilde c}_{\downarrow}(k_0,-{\bm k})|\Psi_{\rm Cooper}\rangle
\label{iF}
\end{equation}
times $\sqrt 2$,
with
${\tilde c}_s(k_0,{\bm k}) = {\cal N}^{-1/2} \sum_i\sum_{\alpha=0,1}
e^{-i(k_0\alpha + {\bm k}\cdot{\bm r}_i)} {\tilde c}_{i,\alpha,s}$.
Here,
$\langle \Psi_{\rm Mott}|$ denotes the critical antiferromagnetic state
of the corresponding Heisenberg model\cite{jpr_10}
at $-J_{0c} = 1.35\, J_1^{\parallel}$. 
(See supplemental Fig. S4.)
The groundstate has $S$ symmetry, as expected, 
but it also alternates
in sign between Cooper pairs at electron Fermi surface pockets
versus Cooper pairs at the emergent hole bands.
(See Fig. \ref{meanfield}b.)
Figure \ref{s+-d+-}
also shows that the (second) excited state
has $D_{x^2-y^2}$ symmetry, as expected,
and that it alternates in sign
in a similar way.
The present exact results therefore provide evidence
for remnant pairing on the emergent hole bands
that lie below the Fermi level at zero 2D momentum.

{\it Discussion and Conclusions.} 
The electronic structure in single-layer FeSe/STO is qualitatively described by
the combination of Figs. \ref{meanfield}b and \ref{exact}a.
For example,
a fit of inelastic neutron scattering data in iron-pnictide superconductors
to the true linear spinwave spectrum Fig. \ref{meanfield}a,
but at the QCP,
yields $J_1^{\parallel}\cong 110$ meV, $J_1^{\perp} = 0$,
and $J_2^{\parallel}\cong 40$ meV $\cong J_2^{\perp}$
for the Heisenberg exchange coupling constants\cite{jpr_10}.
Hopping parameters set in Figs. \ref{meanfield}b and \ref{exact}a
imply that the bottom of the electron bands lies $\epsilon_F \cong 60$ meV
below the Fermi level.
Also,
the cSDW spin gap displayed by Fig. \ref{meanfield}a at sub-critical Hund coupling
is approximately $50$ meV,
which therefore implies that the emergent hole bands
at zero 2D momentum lie $110$ meV below the Fermi.  
Both energy levels are roughly consistent with 
ARPES in single-layer FeSe/STO\cite{liu_12}.
Last, 
the mean-field and exact spectra displayed by Figs. \ref{meanfield}b and \ref{exact}a
predict that ``replicas'' of the $d_{xz}$/$d_{yz}$ buried hole bands
exist at the corner of the one-iron Brillouin zone,
but with orbital quantum numbers interchanged
and at lower energy.
A substrate leads to two inequivalent iron atoms
per hopping of electrons in $d_{xz}$  and $d_{yz}$ orbitals
to neighboring sites.
Zone-folding of
the ``replica'' bands at lower energy to
the center of the two-iron Brillouin zone
possibly
accounts for the ``$D^\prime$ replicas''
of the buried hole bands
that are observed by ARPES on FeSe/STO\cite{lee_14}.


Figure \ref{s+-d+-} predicts $s$-wave Cooper pairs on 
the electron Fermi surface pockets at cSDW momenta.  
This is consistent with
ARPES and with STM on heavily electron-doped surfaces of FeSe,
which find a gap on the electron Fermi surface pockets,
and no evidence for nodes\cite{peng_14,fan_15,zhao_16,yan_15,ye_16,song_16}.
Notably absent from our local moment model (\ref{tJ}) is 
the $3d_{xy}$  electron orbital of the iron atom.
DFT calculations predict inner and outer electron Fermi surface pockets
at the corner of the two-iron Brillouin zone
that have $d_{xz}/d_{yz}$ and $d_{xy}$ orbital character, 
respectively\cite{andersen_boeri_11}.
In such case,
the limit of strong on-site Coulomb repulsion assumed here
would require remnant $s$-wave pairing of opposite sign
on the buried $d_{xy}$ band at the center of the Brillouin zone.
The spectral weight of this band
is negligibly small compared to that of the buried $d_{xz}/d_{yz}$ hole bands
according to high-resolution ARPES on alkali-metal dosed FeSe\cite{ye_16}, however.
This contradiction argues that the iron $3d_{xy}$ orbital 
does not play an important role 
in high-temperature superconductivity shown at surface layers
of heavily electron-doped FeSe.

Figure \ref{s+-d+-} also predicts remnant Cooper pairs of opposite sign on the
emergent hole bands that lie below the Fermi level at zero 2D momentum.
The remnant pairs
are possibly a result of the intrinsic broadening in frequency
experienced by the emergent holes. (Cf. ref. \cite{chen_15}.)
Recent quasi-particle interference patterns obtained from
surface layers of intercalated FeSe
observe a feature at cSDW wavenumbers
that could be accounted for by
the superposition of an electron near cSDW momenta 
with an Andreev reflected hole near zero 2D momentum\cite{yan_15}.
Remnant hole pairing can be confirmed in this way.

{\it Note added:} Recent inelastic neutron scattering studies of intercalated FeSe find low-energy spin resonances in the superconducting state at wavenumbers 
${\bm Q} = (\pi/a)({\bm{\hat x}}+{\bm{\hat y}}) \pm \delta (\pi/a){\bm{\hat x}} ({\bm{\hat y}})$
in the one-iron Brillouin zone\cite{new_ins}, with $\delta = 0.32 - 0.47$.
Comparison of Fig. 1a with supplemental Fig. S1 reveals that
 true spin waves become degenerate with hidden spin waves precisely at such wavenumbers
($\delta = 0.36$).  This observation suggests that hidden magnetic order of the type displayed in the inset to Fig. 1a is present in intercalated FeSe.

\begin{acknowledgments}
The author thanks Nick Bonesteel, Pedro Schlottmann and Oskar Vafek for discussions.
He also thanks
Brent Andersen, Richard Roberts and Timothy Sell for
technical help with the use of
the shared-memory machine (Predator) at
the AFRL DoD Supercomputing Resource Center.
This work was supported in part by the US Air Force
Office of Scientific Research under grant no. FA9550-13-1-0118
and by the National Science Foundation under PREM grant no. DMR-1523588.
\end{acknowledgments}

%

\pagebreak
\widetext
\begin{center}
\textbf{\large Supplemental Material: Isotropic Cooper Pairs with Emergent Sign Changes
 in Single-Layer Iron Superconductor}
\end{center}

\bigskip
\begin{center}
\text{Jose P. Rodriguez}
\end{center}


\medskip
\begin{center}
\it{Department of Physics and Astronomy,}
\end{center}
\begin{center}
\it{California State University at Los Angeles, Los Angeles, CA 90032}
\end{center}

\setcounter{equation}{0}
\setcounter{figure}{0}
\setcounter{table}{0}
\setcounter{page}{1}
\makeatletter
\renewcommand{\theequation}{S\arabic{equation}}
\renewcommand{\thefigure}{S\arabic{figure}}
\renewcommand{\bibnumfmt}[1]{[S#1]}
\renewcommand{\citenumfont}[1]{S#1}

\section{I. Schwinger-Boson-Slave-Fermion Mean Field Theory}
It is first convenient to write the spin-operator in the particle-hole-conjugate form:
${\bm S}_{i,\alpha} =
-(\hbar/2) \sum_{s,s{\prime}}
{\tilde c}_{i,\alpha,s}
{\boldsymbol\sigma}_{s,s^{\prime}} {\tilde c}_{i,\alpha,s^{\prime}}^{\dagger}$.
Substitution of the composite form
for the creation operator of the correlated electron {\it above} half filling,
${\tilde c}_{i,\alpha,s}^{\dagger} = f_{i,\alpha}^{\dagger} b_{i,\alpha,s}$,
then yields the expression
${\bm S}_{i,\alpha} =
-(\hbar/2) \sum_{s,s{\prime}}
f_{i,\alpha} b_{i,\alpha,s}^{\dagger}
{\boldsymbol\sigma}_{s,s^{\prime}} b_{i,\alpha,s^\prime} f_{i,\alpha}^{\dagger}$.
Next, replacing the operator $f_{i,\alpha} f_{i,\alpha}^{\dagger}$
with its expectation value, $1-x$, yields the approximation
\begin{equation}
{\bm S}_{i,\alpha} \cong
-(1-x) {1\over 2} \hbar \sum_{s,s{\prime}}
b_{i,\alpha,s}^{\dagger}
{\boldsymbol\sigma}_{s,s^{\prime}} b_{i,\alpha,s^\prime}
\label{s_mf_spin_operator}
\end{equation}
for the spin operator.
Here, $x$ denotes the concentration of mobile electrons per orbital.
Last,
we shall also neglect on-site repulsion $U_0^{\prime}$
between a mobile electron in the $d-$ orbital and  
a mobile electron in the $d+$ orbital.
This approximation should be valid in the dilute limit, $x\rightarrow 0$.
As mentioned in the paper, it is also convenient
to next rotate the spins quantized along the $z$ axis
by an angle $\pi$ about the $y$ axis
on one of the antiferromagnetic sublattices
in the hidden magnetic order shown by the inset to Fig. 1a in the paper; e.g.,
$b_{i,\beta,\uparrow}^{\dagger} \rightarrow - b_{i,\beta,\downarrow}^{\dagger}$ and
$b_{i,\beta,\downarrow}^{\dagger} \rightarrow b_{i,\beta,\uparrow}^{\dagger}$,
for $(i,\beta)$ that lie in the down-spin sublattice.
This decouples spins between the two hidden antiferromagnetic
sublattices\cite{s_Arovas_Auerbach_88}.

Let us now turn off nearest-neighbor intra-orbital hopping in 
the two-orbital $t$-$J$ model, Eq. (1) in the paper:
$t_1^{\parallel} = 0$.
Mean fields among the Schwinger bosons are pair amplitudes
across the antiferromagnetic links\cite{s_Arovas_Auerbach_88}:
$Q_0 = \langle b_{i,d-,s} b_{i,d+,s}\rangle$,
$Q_{1}^{\parallel} = \langle b_{i,d\pm,s} b_{j,d\pm,s}\rangle$ and
$Q_{2}^{\perp} = \langle b_{i,d\pm,s} b_{j,d\mp,s}\rangle$.
Here, the superscripts
$\parallel$ and $\perp$ denote intra-orbital ($d\pm$ $d\pm$)
and inter-orbital ($d\pm$ $d\mp$) links,
while the subscripts $0$, $1$ and $2$ denote on-site, nearest neighbor and
next-nearest neighbor links.
On the other hand,
mean fields among the Schwinger bosons
are hopping amplitudes
across the ferromagnetic links\cite{s_Arovas_Auerbach_88}:
$Q_{1}^{\perp} = \langle b_{i,d\pm,s}^{\dagger} b_{j,d\mp,s}\rangle$ and
$Q_{2}^{\parallel} = \langle b_{i,d\pm,s}^{\dagger} b_{j,d\pm,s}\rangle$.
Last,
nearest-neighbor hopping of electrons across the two orbitals
is accounted for by the mean field
among slave fermions
$P_1^{\perp} = {1\over 2} \langle f_{i,d\pm}^{\dagger} f_{j,d\mp}\rangle$,
which has $d$-wave symmetry:
$P_1^{\perp} ({\bm{\hat y}}) = - P_1^{\perp} ({\bm{\hat x}})$.
The dynamics of free Schwinger bosons is then governed by the Hamiltonian 
$$H_b = {1\over 2}\sum_{k}\sum_{s} \{
\Omega_{fm}(k) [b_s^{\dagger}(k) b_s(k) +  b_s(-k) b_s^{\dagger}(-k)]
            + \Omega_{afm}(k) [b_s^{\dagger}(k) b_s^{\dagger}(-k) + b_s(-k) b_s(k)]\},$$
with diagonal and off-diagonal matrix elements
%
\begin{eqnarray*}
\Omega_{fm}(k) &=&
\delta\lambda + J_0^{\prime} Q_0 +
4 J_1^{\prime\parallel} Q_1^{\parallel} + 4 J_2^{\prime\perp} Q_2^{\perp}\\
&& - 4 [J_1^{\prime\perp} Q_1^{\perp} - 2 t_{1}^{\perp}({\bm{\hat x}}) P_{1}^{\perp}({\bm{\hat x}})][1-e^{ik_0}\gamma_{1+}({\bm k})]
- 4 J_2^{\prime\parallel} Q_2^{\parallel}[1-\gamma_2({\bm k})]\\
\Omega_{afm}(k) &=&
-J_0^{\prime} Q_0 e^{i k_0}
-4 J_1^{\prime\parallel} Q_1^{\parallel} \gamma_{1+}({\bm k})
-4 J_2^{\prime\perp} Q_2^{\perp}e^{ik_0}\gamma_2({\bm k}) ,
\end{eqnarray*}
while the dynamics of free slave fermions is then governed by the Hamiltonian 
$H_f = \sum_k \varepsilon_f(k) f^{\dagger}(k) f(k)$,
with the energy eigenvalues
$$\varepsilon_f(k) = -8t_{1}^{\perp}({\bm{\hat x}}) Q_1^{\perp} e^{ik_0} \gamma_{1-}({\bm k}).$$
From here on we set $\hbar = 1$.
Above,
$k = (k_0,{\bm k})$
is the  3-momentum for these excitations,
with corresponding destruction operators
$b_s(k) = {\cal N}^{-1/2}\sum_{\alpha = 0}^1\sum_i e^{-i(k_0 \alpha + {\bm k}\cdot{\bm r}_i)} b_{i,\alpha,s}$ and
$f(k) = {\cal N}^{-1/2}\sum_{\alpha = 0}^1\sum_i e^{-i(k_0 \alpha + {\bm k}\cdot{\bm r}_i)} f_{i,\alpha}$.
Here,
${\cal N} = 2 N_{\rm Fe}$ denotes the number of site-orbitals
on the square lattice of $N_{\rm Fe}$ iron atoms,
while the indices $0$ and $1$ denote
the $d-$ and $d+$ orbitals $\alpha$.
The quantum numbers $k_0 = 0$ and $\pi$ therefore represent
the $d_{xz}$ and the $(-i)d_{yz}$ orbitals.
Also above,
$\gamma_{1\pm} ({\bm k}) = {1\over 2} ({\rm cos}\, k_x a\, \pm \,{\rm cos}\, k_y a)$
and $\gamma_2 ({\bm k}) = {1\over 2} ({\rm cos}\, k_+ a\,+\,{\rm cos}\, k_- a)$,
with $k_{\pm} = k_x \pm k_y$.
The infinite-$U_0$ constraint,
Eq. (2) in the paper,
is enforced on {\it average} over the bulk of the system
by the boson chemical potential, $\delta\lambda$,
while
the chemical potential of the slave fermions, $\mu$,
sets the concentration of mobile  electrons per site-orbital, $x$.
Last, the mean-field approximation (\ref{s_mf_spin_operator})
that accounts for the effect of mobile electrons on the spin operator 
results in 
effective Heisenberg spin-exchange coupling constants\cite{s_Auerbach_Larson_91}
$J^{\prime} = (1-x)^2 J$.

\begin{figure}
\includegraphics[scale=0.65, angle=-90]{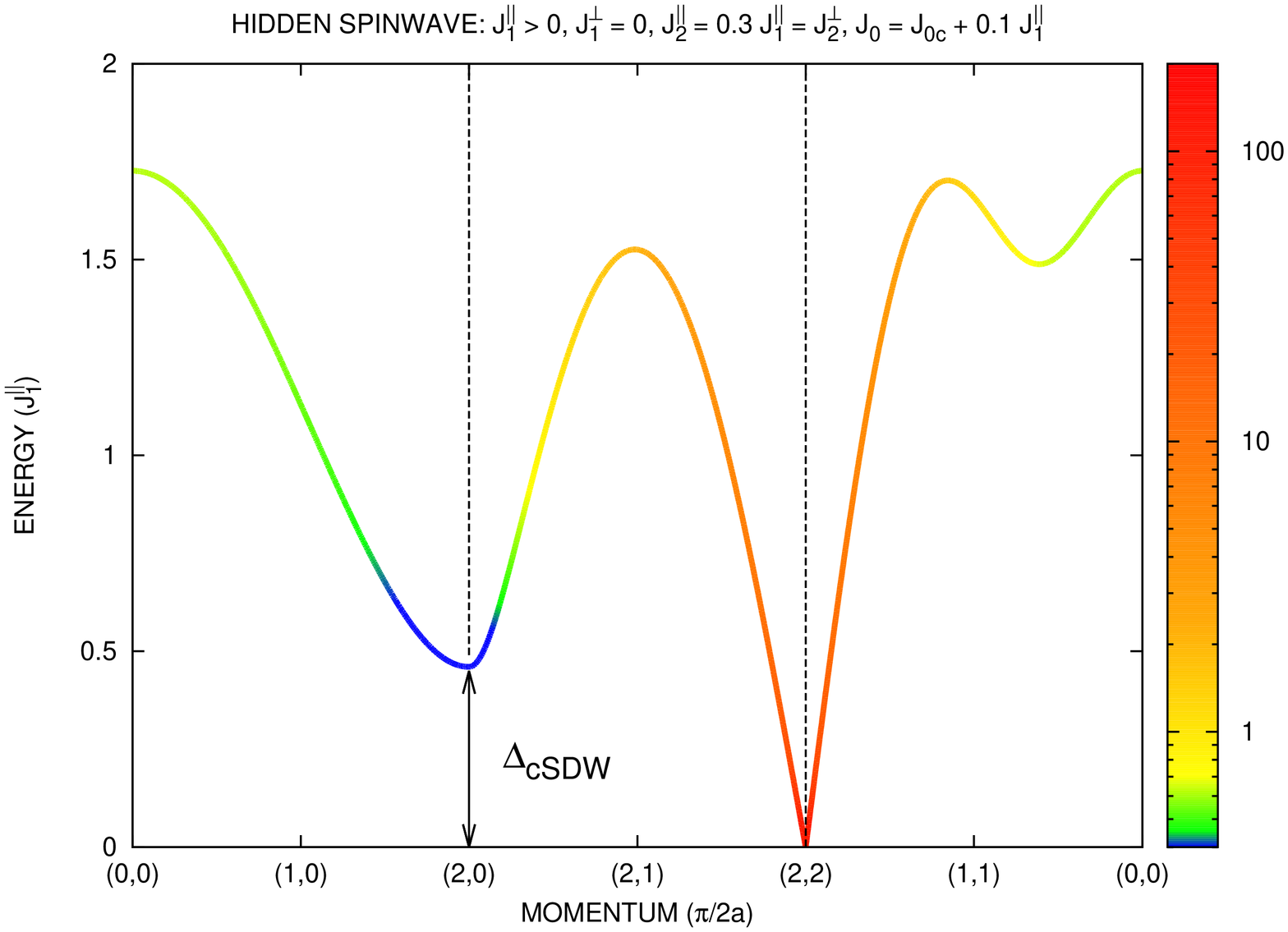}
\caption{The imaginary part of the transverse spin susceptibility,
Eq. (3) in the paper, at site-orbital concentration $x=0.01$,
in the hidden spin channel.
Hopping matrix elements
are ideal: $t_1^{\parallel} = 0$, 
$t_1^{\perp}({\bm{\hat x}}) = + 5 \, J_1^{\parallel}$, and
$t_1^{\perp}({\bm{\hat y}}) = - 5 \, J_1^{\parallel}$.}
\label{s_meanfield_b}
\end{figure}

The solution to the above mean field theory is achieved by making
the standard Bogoliubov transformation
of the boson field\cite{s_Arovas_Auerbach_88}:
$b_s(k) = ({\rm cosh}\, \theta_k) \beta_s(k) + ({\rm sinh}\, \theta_k) \beta_s^{\dagger}(-k)$,
 with
${\rm cosh} \, 2\theta = \Omega_{fm} / \omega_b$ and
${\rm sinh} \, 2\theta = -\Omega_{afm} / \omega_b$, where
$\omega_b = (\Omega_{fm}^2 - \Omega_{afm}^2)^{1/2}$
is the energy of the boson ($\beta$).
Enforcing the infinite-$U_0$ constraint 
[Eq. (2) in the paper]
on average then  results in
ideal Bose-Einstein condensation (BEC) of the Schwinger bosons
into degenerate groundstates at
$k = 0$ and $(\pi,\pi/a,\pi/a)$
as temperature $T\rightarrow 0$,
in which case $\delta\lambda\rightarrow 0$.
(See paper, Fig. 1a, and see Fig. \ref{s_meanfield_b}.)
All five mean fields among the Schwinger bosons take on the unique value $Q = s_0$
at large-$s_0$ under ideal BEC\cite{s_jpr_mana_pds_11,s_jpr_mana_pds_14}.
Slave fermions
in $d_{xz}$ and $d_{yz}$ orbitals
condense inside of
circular Fermi surfaces centered at wavenumbers
$(\pi/a){\bm{\hat y}}$ and $(\pi/a){\bm{\hat x}}$, respectively,
at low electron doping $x \ll 1$, with Fermi wave vector $k_F a = (4\pi x)^{1/2}$.
(See the inset to Fig. 1b in the paper.)
The mean inter-orbital electron hopping amplitude is then approximately
$P_{1}^{\perp}({\bm{\hat x}}) = x/2$.

Equation (3) in the paper for
the dynamical spin correlation function
$\langle S^{(+)} S^{\prime(-)}\rangle |_{k,\omega}$
of the hidden N\'eel half metal
is a direct application of the Auerbach-Arovas expression for
the auto-correlation function
$\langle S_y S_y^{\prime}\rangle |_{k,\omega}$
at ideal BEC of the Schwinger bosons\cite{s_jpr_mana_pds_14,s_Auerbach_Arovas_88},
multiplied by a factor of two because of spin isotropy.
The result notably coincides with that obtained
within the linear spin-wave approximation 
at the large-$s_0$ limit\cite{s_jpr_10}.
Figure \ref{s_meanfield_b} gives the hidden-order counterpart 
to the spectrum of true spinwaves near the QCP
predicted by this mean-field approximation,
Fig. 1a in the paper.  
As expected by general considerations\cite{s_jpr_10},
these spectra are shifted with respect to each other
by momentum $(\pi / a)({\bm{\hat x}}+{\bm{\hat y}})$.

Also, within the above mean field theory,
the one-electron propagator
is given by 
the convolution
of the propagator for slave fermions ($f$)
with the conjugate propagator for Schwinger bosons ($b$) 
in 3-momentum and in frequency:
$i G(k,\omega) = G_b^* * G_f |_{k,\omega}$.
Here, the propagator for free slave fermions reads
$G_f (k,\omega) = [\omega + \mu - \varepsilon_f (k)]^{-1}$,
while  the propagator for free Schwinger bosons reads
$G_b (k,\omega) = ({\rm cosh} \, \theta_k)^2[\omega - \omega_b(k)]^{-1}
-({\rm sinh} \, \theta_k)^2 [\omega + \omega_b(k)]^{-1}$.
The conjugate propagator for free Schwinger bosons evolving backwards in time is then
$G_b^* (k,\omega) = -({\rm cosh} \, \theta_k)^2[\omega + \omega_b(k)]^{-1}
+({\rm sinh} \, \theta_k)^2 [\omega - \omega_b(k)]^{-1}$.
After rewriting the resulting products of poles as sums/differences of poles,
standard summations of Matsubara frequencies  
yield expression (4) given in the paper.
There, the identities
$({\rm cosh}\,\theta)^2 = {1\over 2} {\rm cosh}\,2\theta + {1\over 2}$
 and
$({\rm sinh}\,\theta)^2 = {1\over 2} {\rm cosh}\,2\theta - {1\over 2}$ 
have been used.

\section{II. Exact Diagonalization}
Hidden magnetic order of 
the type depicted by the inset to Fig. 1a in the paper
is predicted by  the two-orbital Heisenberg model over the square lattice
in the large-$s_0$ limit
for exhange coupling constants that exhibit diagonal frustration,
at weak to moderate Hund coupling\cite{s_jpr_10,s_jpr_ehr_09}:
e.g., $J_1^{\parallel} > 0$, $J_1^{\perp} = 0$, 
$J_2^{\parallel} = 0.3\, J_1^{\parallel} = J_2^{\perp}$,
and $J_0 = J_{0c} + 0.1\, J_1^{\parallel}$,
where $-J_{0c}$ is the quantum-critical Hund coupling at which
the spin gap associated with commensurate spin-density wave (cSDW)
order collapses to zero.
Add now electrons with ideal nearest-neighbor hopping; e.g.,
$t_1^{\parallel} = 0$,
$t_1^{\perp}({\bm{\hat x}}) = +5\, J_1^{\parallel}$ and
$t_1^{\perp}({\bm{\hat y}}) = -5\, J_1^{\parallel}$. 
Spin-polarized electrons hop within each antiferromagnetic sublattice in such case.
The critical Hund coupling is given by
$- J_{0c} = 2 (J_1^{\parallel} - J_1^{\perp}) - 4 J_2^{\parallel}
  + (1-x)^{-2} s_0^{-1} 2  t_{1}^{\perp}({\bm{\hat x}}) x$
within the mean-field approximation.  
Adding mobile electrons thereby
stabilizes the hidden N\'eel order.
Figure \ref{s_meanfield_b} reveals the Goldstone mode at N\'eel wavenumber
$(\pi/a)({\bm{\hat x}}+{\bm{\hat y}})$
expected from such hidden antiferromagnetic order 
for the corresponding half metal state of the two-orbital $t$-$J$ model
within the mean-field approximation\cite{s_jpr_mana_pds_11,s_jpr_mana_pds_14}.
It appears as a divergence in the imaginary part
of the transverse susceptibility, Eq. (3) in the paper, for hidden spin,
${\bm S}_{i,d-} - {\bm S}_{i,d+}$.
Recall that spin-$1/2$ moments live on the
$d\pm = d_{(x\pm iy)z}$ orbitals.
The Schwinger-boson-slave-fermion mean-field approximation
for the two-orbital $t$-$J$ model
employed in the paper
also predicts coherent electron bands that result in Fermi surface pockets
centered at cSDW momenta
$(\pi/a){\bm{\hat{x}}}$ and $(\pi/a){\bm{\hat{y}}}$.
(See the inset to Fig. 1b in the paper.)
Below, we compare this mean field theory to
exact results in the absence of Hund's Rule, $J_0 = 0$,
where the hidden half metal state is most stable.

\begin{figure}
\includegraphics[scale=0.65, angle=-90]{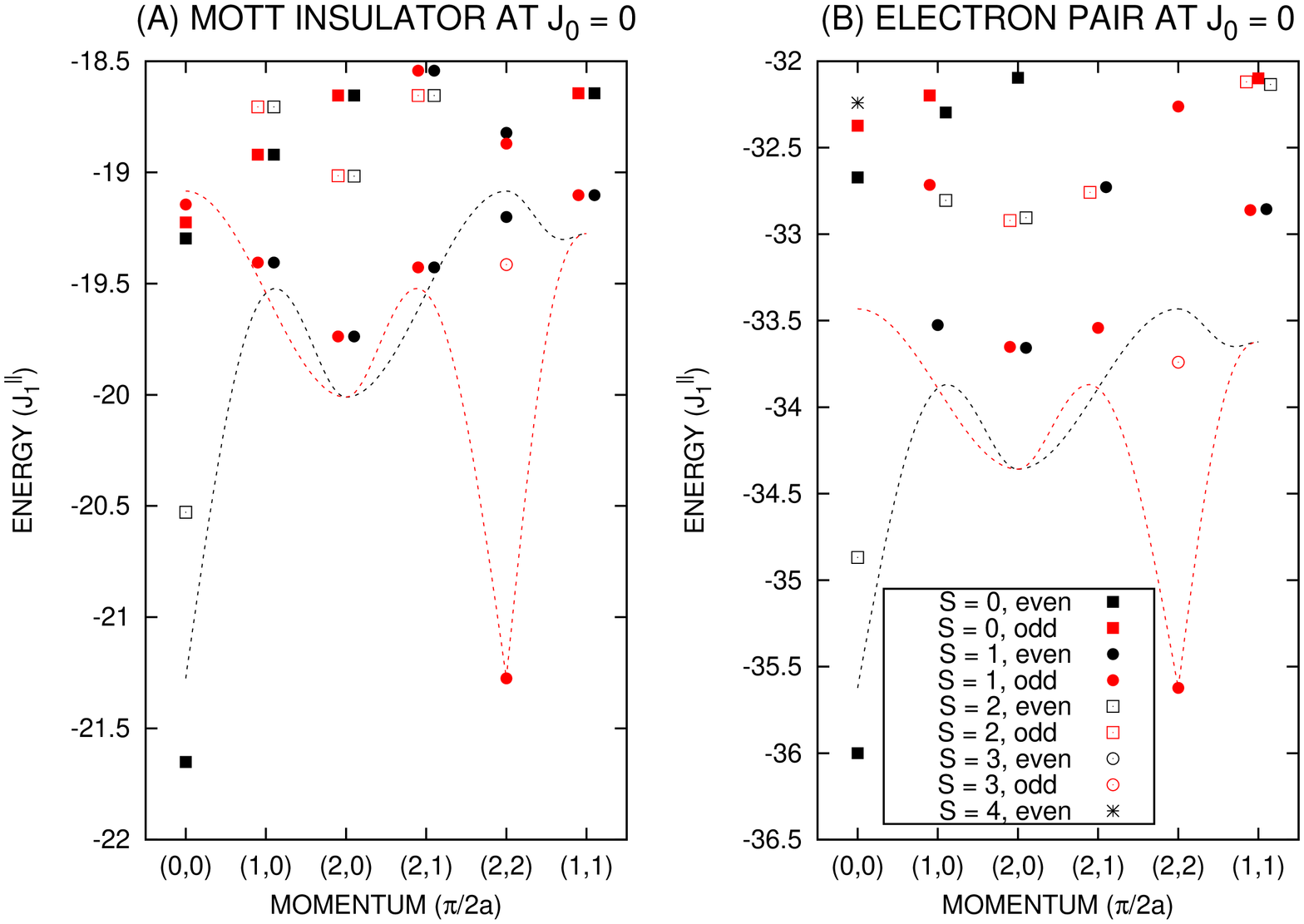}
\caption{(a) Exact spectra of two-orbital Heisenberg model over a
periodic $4\times 4$ lattice,
with exchange coupling constants
that coincide with those in Fig. \ref{s_meanfield_b}.
Black states are even under orbital exchange,
$P_{d,{\bar d}}$, while red states are odd under it.
Henceforth, some points in spectra are artificially
moved slightly off their quantized values along the momentum axis
for the sake of clarity.
(b) Exact spectrum of two electrons more than half filling for the corresponding
two-orbital $t$-$J$ model
plus repulsive interactions (see paper),
plus constant ${1\over 4}(N_{\rm Fe}-2)J_0$.
Hopping matrix elements coincide with those in Fig. \ref{s_exact_31},
and $U_0^{\prime} = {1\over 4}J_0 + 1000\, J_1^{\parallel}$.
In both panels,
red and black dashed lines
trace the dispersion of hidden 
and of true spinwaves, respectively,
in the limit $x\rightarrow 0$
following Eq. (3) in the paper.
}
\label{s_exact_32_30}
\end{figure}

{\it Hund's Rule Absent.}
Figure \ref{s_exact_32_30}a compares
exact results for the low-energy spectrum of the frustrated Heisenberg
model on a periodic $4\times 4$ lattice
of iron atoms with $d-$ and $d+$ orbitals
to
the spin-wave spectrum predicted by Schwinger-boson mean field theory
for hidden magnetic order,
Eq. (3) in the paper.
Heisenberg exchange coupling constants are set by Fig. \ref{s_meanfield_b},
but without Hund's Rule: $J_0 = 0$.  Also, the concentration of mobile electrons
per site-orbital is set to $x=0$ in all mean-field expressions.
Black states in Fig. \ref{s_exact_32_30}a have
even parity under orbital exchange $P_{d,{\bar d}}$,
while red states have odd parity under it.
Black spin-$1$ states therefore
represent true spin fluctuations,
while red spin-$1$ states  represent hidden spin fluctuations.
Notice that the predicted spin-wave spectrum for the hidden N\'eel state
traced by the dashed lines in Fig. \ref{s_exact_32_30}a
successfully describes the dispersion of the exact spin-$1$ states 
at low energy\footnote{The spectral weight of true (``black'')
spinwaves at zero 2D momentum
is identically zero [cf. Eq. (3) and Fig. 1a in the paper],
hence the absence of spin-$1$ states there 
in Figs. \ref{s_exact_32_30}a and \ref{s_exact_32_30}b.}.
Notice also the tower in Fig. \ref{s_exact_32_30}a beginning with the
spin-$0$ groundstate at zero 2D momentum,
the spin-$1$ first-excited state at 2D momentum
$(\pi / a)({\bm{\hat x}} + {\bm{\hat y}})$,
the spin-$2$ second-excited state back at zero 2D momentum,
and the spin-$3$ excited state back at 2D momentum 
$(\pi / a)({\bm{\hat x}} + {\bm{\hat y}})$.
This tower of spin-$n$ states clearly coincides with multiply-occupied states
of the hidden order spinwave, which is occupied $n$ times.

\begin{figure}
\includegraphics[scale=0.65, angle=-90]{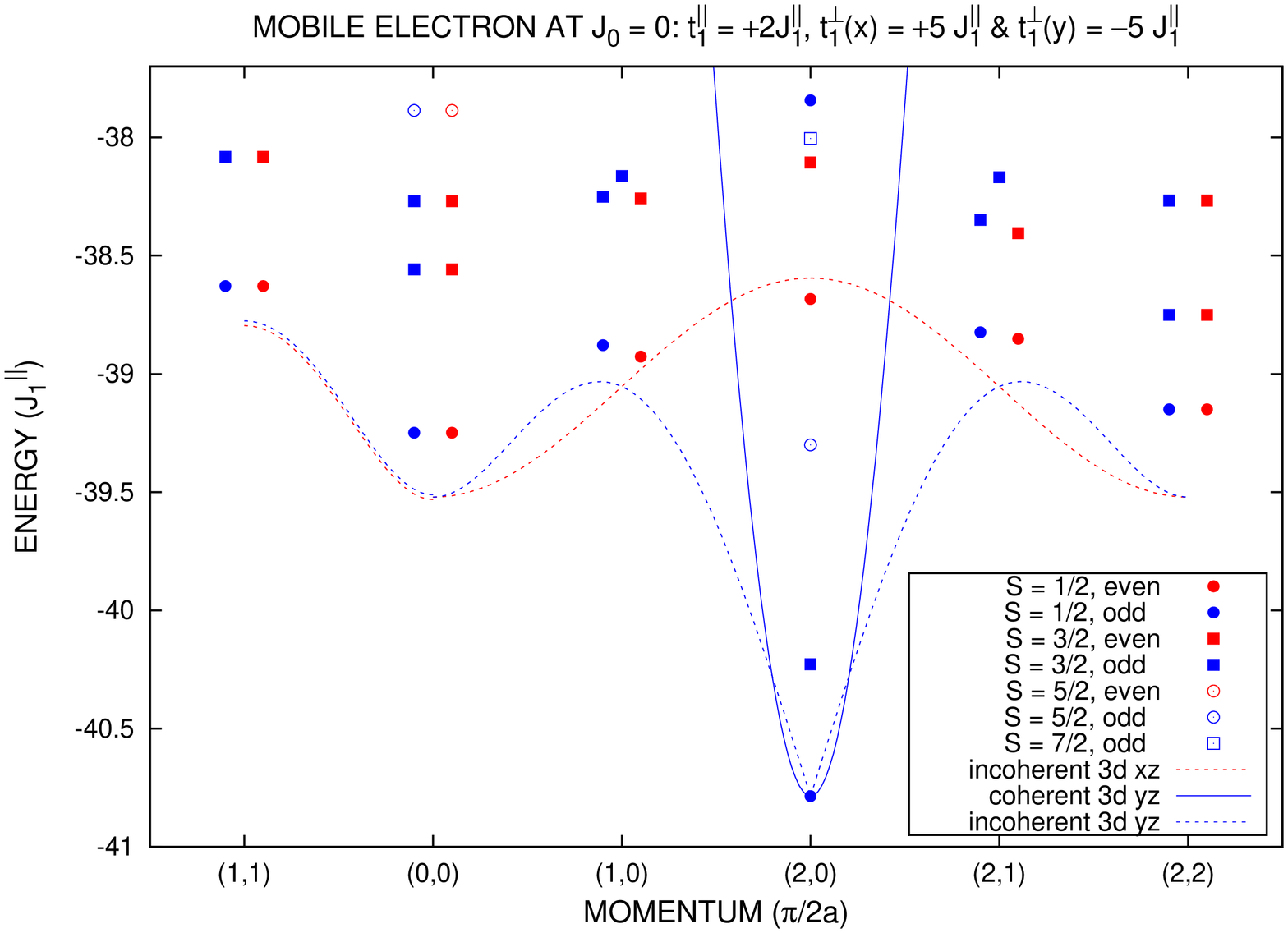}
\caption{
The exact spectrum of one electron more than half filling for the
two-orbital $t$-$J$ model over a $4\times 4$ periodic lattice
in the absence of Hund's Rule. Heisenberg exchange coupling constants coincide
with those in Fig. \ref{s_meanfield_b}. 
Red and blue dashed lines trace
the dispersion of emergent hole excitations 
in  $d_{xz}$ and $d_{yz}$ orbitals, respectively,
at ideal electron hopping, $t_1^{\parallel} = 0$,
as predicted by Eq. (5) in the paper.}
\label{s_exact_31}
\end{figure}

Figure \ref{s_exact_31} shows the exact low-energy spectrum of one electron
more than half filling governed by the two-orbital $t$-$J$ model,
Eq. (1) in the paper, in the absence of Hund's Rule.
Heisenberg exchange coupling constants coincide with
those in Figs. \ref{s_meanfield_b} and \ref{s_exact_32_30},
while hopping matrix elements are set to
$t_1^{\parallel} = 2\, J_1^{\parallel}$,
$t_{1}^{\perp}({\bm{\hat x}}) = +5\, J_1^{\parallel}$ and
$t_{1}^{\perp}({\bm{\hat y}}) = -5\, J_1^{\parallel}$.
Red states are even under orbital exchange $P_{d,{\bar d}}$,
while blue states are odd under it.
The solid blue line depicts the $d_{yz}$ half metal band predicted by
Schwinger-boson-slave-fermion mean field theory at the limit
towards half filling, $x \rightarrow 0$,
but with ideal electron hopping, $t_1^{\parallel} = 0$.
The dashed lines trace the dispersion of emergent hole excitations
predicted by Eq. (5) and Fig. 1b of the paper.
They successfully describe the dispersion of the exact spin-$1/2$ groundstates
per orbital quantum number in the absence of Hund's Rule.  
Notice, however, the first-excited states per momentum
that carry spin $3/2$ in Fig. \ref{s_exact_31}.
The pairs of spin-$1/2$ and spin-$3/2$ states that they make up
per momentum
can be understood as the result of the addition of angular momentum
between a spin-$1/2$ electron at cSDW wavenumbers and
a spin-$1$ spinwave in the half metal\cite{s_jpr_mana_pds_14}.
In particular, 
the spin-$3/2$ state at momentum $(\pi/a){\bm{\hat x}}$ with $d_{yz}$-orbital symmetry
shown in Fig. \ref{s_exact_31}
can be understood as a spin-$1/2$ electron in orbital $d_{xz}$ at momentum
$(\pi/a){\bm{\hat y}}$ combined with a hidden-order (odd-parity)  spinwave
that carries momentum $(\pi/a)({\bm{\hat x}}+{\bm{\hat y}})$.
In turn, the second-excited spin-$5/2$ state at this momentum,
which has the same $d_{yz}$-orbital symmetry,
can be understood as the combination of the spin-$1/2$ groundstate
with {\it two} hidden-order spin-waves.  This tower of states resembles
the previous one identified at half filling in Fig. \ref{s_exact_32_30}a.

\begin{figure}
\includegraphics[scale=0.65, angle=-90]{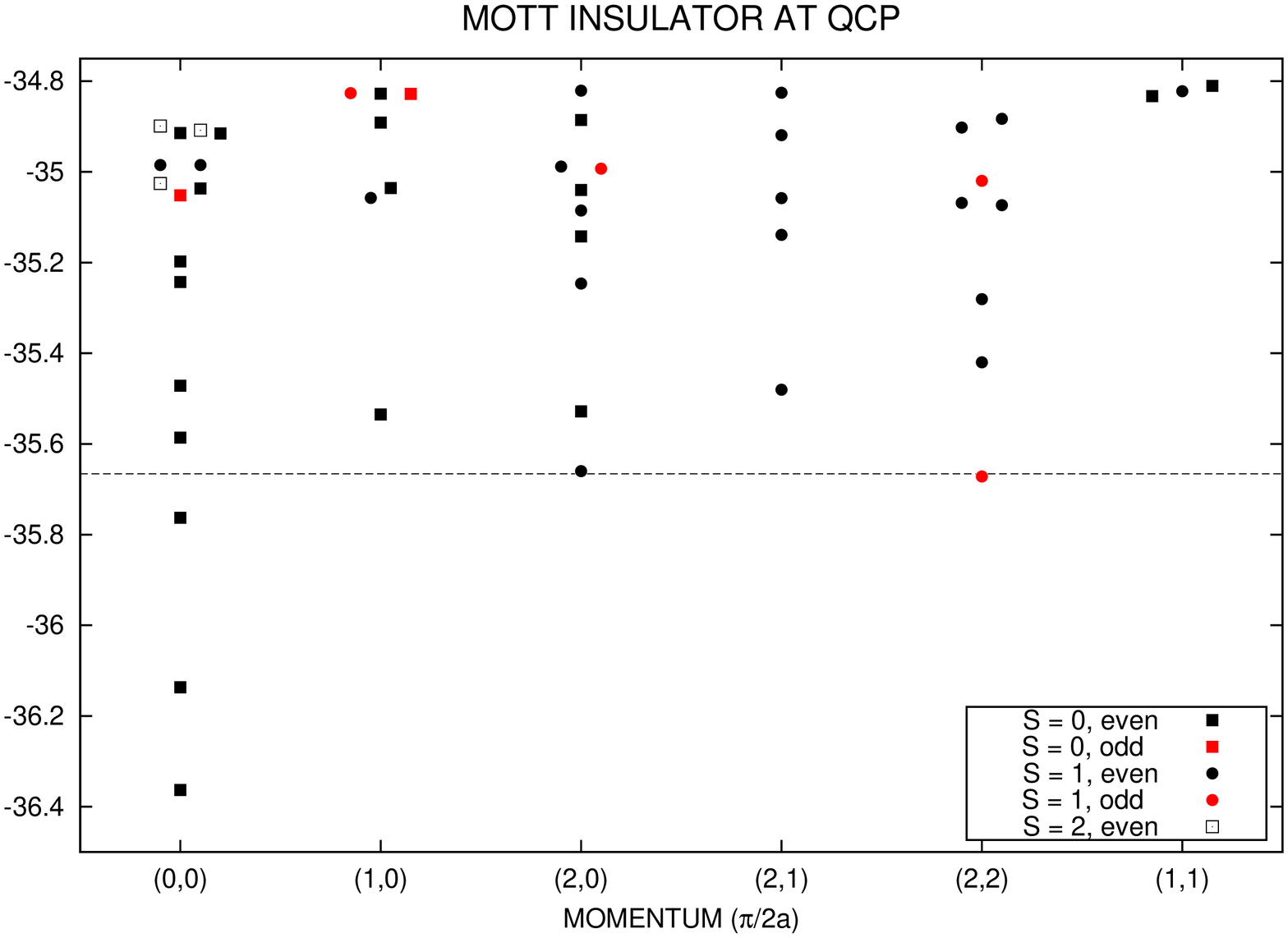}
\caption{Heisenberg exchange coupling constants 
coincide with Fig. \ref{s_meanfield_b},
but the Hund coupling is tuned to
the critical value $-J_0 = 1.35 \,J_1^{\parallel}$.}
\label{s_exact_32_qcp}
\end{figure}

Last,
Fig. \ref{s_exact_32_30}b compares the exact spectrum of
two electrons more than half-filling in the absence of Hund's Rule with
the spin-excitation spectrum predicted by Schwinger-boson-slave-fermion mean field theory at ideal electron  hopping $t_1^{\parallel} = 0$.
Notice that the tower of $n=0,1,2$ and $3$-occupied
hidden-order spinwave states persists$^1$.  
Comparison with Fig. \ref{s_exact_32_30}a
indicates that a  gap separates out the tower of lowest-energy states
in the case of two mobile electrons.

We therefore
conclude that Schwinger-boson-slave-fermion mean field theory
is a valid approximation for
the two-orbital $t$-$J$ model
in the case of the hidden half metal state
depicted by the insets to Fig. 1 of the paper. 
(Cf. refs. \cite{s_jpr_mana_pds_11} and \cite{s_jpr_mana_pds_14}.)
In particular, in the absence of Hund's Rule, 
Fig. \ref{s_exact_32_30}
demonstrates that it works well for spin-$1$ states
at both half filling and in the case of two mobile electrons.
Again, in the absence of Hund's Rule,
Fig. \ref{s_exact_31} demonstrates that
Schwinger-boson-slave-fermion mean field theory also works well 
in the case of one mobile electron for spin-$1/2$ states.

{\it QCP.}
Figure \ref{s_exact_32_qcp} shows the exact spectrum of
the same two-orbital Heisenberg model 
that corresponds to Fig. \ref{s_exact_32_30}a,
but at the putative quantum-critical point.  
Here, the Hund coupling is tuned to
the critical value $-J_0 = 1.35\, J_1^{\parallel}$ at which
the lowest energy spin-$1$ states 
at cSDW wavenumbers $(\pi/a){\bm{\hat x}}$ and $(\pi/a){\bm{\hat y}}$
become degenerate with 
the lowest-energy spin-$1$ state
at N\'eel wavenumber $(\pi/a)({\bm{\hat x}}+{\bm{\hat y}})$.
The former states 
are true spin fluctuations,
with even parity under $P_{d,{\bar d}}$,
while the latter state 
is a hidden spin fluctuation,
with odd parity under $P_{d,{\bar d}}$.


\begin{thebibliography}{}
\bibitem{diverge} At wavenumbers ${\bm k} = 0$ and $(\pi/a)({\bm{\hat x}}+{\bm{\hat y}})$,
the spectral weight in (\ref{emergent_holes}) diverges
as the cSDW spin gap vanishes at the QCP.
It yields  the bound
$\Delta_{cSDW} > k_F v_0$
that  guarantees the
validity of (\ref{emergent_holes}).
Here, $v_0$ coincides with the velocity of cSDW spinwaves at the QCP\cite{jpr_10}.
The bound $|{\bm k}-{\bm Q}_{q_0}| > k_F$,
likewise,
guarantees the validity of (\ref{emergent_holes})
at ${\bm k}$ near cSDW wavenumbers ${\bm Q}_{q_0}$.
\bibitem{broadening} At $t_1^{\parallel} = 0$, the convolution (\ref{G})
implies intrinsic broadening of the electron bands below the Fermi level:
$\Delta\omega \sim {1\over 8}(\epsilon_F/2 s_0) 
= {1\over 2}\pi x t_1^{\perp}({\bm{\hat x}})$ at $t \gg J$.  
The Goldstone mode associated with hidden magnetic order
(inset to Fig. \ref{meanfield}a)
may no longer exist
once $t_1^{\parallel} \neq 0$, however. 
In such case, 
the coherent and incoherent $d_{yz}$ bands shown in Fig. \ref{meanfield}b
may experience level repulsion at
cSDW momentum $(\pi/a){\bm{\hat x}}$,
likewise the $d_{xz}$ bands at cSDW momentum $(\pi/a){\bm{\hat y}}$.
\bibitem{t1_parallel} In the independent electron approximation,
adding small but positive intra-orbital nearest-neighbor hopping,
$0 < t_1^{\parallel} < t_1^{\perp}({\bm{\hat x}})$,
results in an electron band centered at wavenumber $(\pi/a){\bm{\hat x}}$
for the $d_{yz}$ orbital with mass anisotropy $m_x > m_y$.
The mass anisotropy of the corresponding
emergent hole band
at zero 2D momentum is the reverse 
within the Schwinger-boson-slave-fermion mean-field approximation,
at  ideal hopping, $t_1^{\parallel} = 0$.
Adding the mean fields
$R_1^{\parallel} = \langle b_{i,d\pm,s}^{\dagger} b_{j,d\pm,s}\rangle$ and
$P_1^{\parallel} = {1\over 2}\langle f_{i,d\pm}^{\dagger} f_{j,d\pm}\rangle$
for intra-orbital nearest-neighbor hopping of electrons  
(before any $90$ degree spin rotations about the $y$ axis)
results in elliptical electron Fermi surface pockets at wavenumbers
$(\pi/a){\bm{\hat x}}$ and $(\pi/a){\bm{\hat y}}$
for the $d_{yz}$ and the $d_{xz}$ orbitals, respectively,
with the major axes along the principal axes.
Both of the extra mean fields vanish in the optimized 
theory at large $s_0$, however. (Cf. ref. \cite{jpr_mana_pds_14}.)
\end{thebibliography}
\end{document}